\documentstyle[prl,aps,multicol,epsfig]{revtex}
\renewcommand{\narrowtext}{\begin{multicols}{2} \global\columnwidth20.5pc}
\renewcommand{\widetext}{\end{multicols} \global\columnwidth42.5pc} 

\begin{document}

\newcommand{\be}{\begin{equation}}
\newcommand{\ee}{\end{equation}}
\newcommand{\bea}{\begin{eqnarray}}
\newcommand{\eea}{\end{eqnarray}}
\newcommand{\nt}{\narrowtext}
\newcommand{\wt}{\widetext}

\title{Excitonic instability in two-dimensional degenerate semimetals}
\author{H. Leal and D. V. Khveshchenko}
\address{Department of Physics and Astronomy, University of North
Carolina, Chapel Hill, NC 27599}
\maketitle

\begin{abstract}
We study the possibility of excitonic pairing 
in layered degenerate semimetals such as graphite,
where the electron density of states almost vanishes at the Fermi level
and, therefore, the Coulomb interactions remain essentially unscreened.
By focusing on the Dirac-like low-energy electron excitations 
and numerically solving a non-linear gap equation for the order parameter,
we obtain a critical value of the Coulomb coupling
and establish the Kosterlitz-Thouless-like nature of a
putative semimetal-to-excitonic insulator transition.
\end{abstract}

\nt
In layered graphite, 
a poor screening of the Coulomb interaction sets the stage for a novel form 
of excitonic instability 
resulting in the opening of a gap in the quasi-two-dimensional electronic spectrum 
and manifesting itself through the onset of an insulating charge density wave
\cite{DVK1}.

It was also argued in Ref.\cite{DVK1} that the pseudo-relativistic kinematics 
of the Dirac-like electronic excitations in a single sheet of graphite
allows one to draw a formal parallel with the phenomenon of chiral symmetry breaking (CSB)
that has long been studied in the context of the three-dimensional 
Quantum Electrodynamics ($QED_{2+1}$) and other relativistic fermion theories. 

In the case of $QED_{2+1}$, a zero
temperature CSB transition was predicted to occur for a sufficiently 
small ($N<N_c$) number of fermion
species, regardless of the interaction strength \cite{Appelquist88}, while 
the short-ranged Higgs-Yukawa (HY) four-fermion interactions can drive this transition 
for any $N$, provided that the HY coupling is strong enough \cite{HY}.

The CSB scenario has recently become a common theme of several 
$QED_{2+1}$-like descriptions of the pseudogap phase in
underdoped cuprates \cite{RW,FT,Herbut}. In this regard, some 
authors \cite{FT} cited the original estimates of the 
critical number of fermions $N^{(0)}_c\approx 3.24$ \cite{Appelquist88}
to support the conjecture
that the inherent spin density wave instability of the $d$-wave symmetrical pseudogap state
can be readily described by the conventional 
$QED_{2+1}$ theory with $N=2$ species of
the Dirac-like nodal quasiparticles.

However, the mounting analytical \cite{Appelquist99}
and numerical \cite{Hands} evidence indicating that 
in the conventional $QED_{2+1}$ the actual critical number of flavors
$N_c$ may be less than two seems to suggest otherwise and calls for a need
to further modify the minimal $QED_{2+1}$ theory of the pseudogap phase 
(possibly, beyond recognition).

The implications of the results of Refs.\cite{Appelquist99,Hands}
made other groups \cite{Herbut} emphasize a potential importance 
of additional four-fermion couplings and/or anisotropic quasiparticle dispersion
(in high-$T_c$ cuprates, the quasiparticle velocity is strongly dependent on the direction,
$v_1/v_2\sim 10-20$), whose systematic account has yet to be
done. It has been recently argued, however, that $N_c$
may not be affected by the dispersion anisotropy at all \cite{Lee}, thus 
implying that the four-fermion couplings (which are
absent in the conventional $QED_{2+1}$) may indeed play a crucial
role in the theory of the 
pseudogap-to-antiferromagnet transition in underdoped cuprates. 

In the problem of the conjectured excitonic instability
in graphite, the non-relativistic nature of the Coulomb 
interaction further invalidates any naive attempts to make use of  
the results pertaining solely to the relativistically invariant systems.
Nevertheless, the earlier work on the subject has already provided for some analytical
evidence
that the excitonic transition may indeed occur for 
a sufficiently strong Coulomb coupling \cite{DVK1,Gorbar}. 

It was also predicted \cite{DVK2} that a magnetic field normal to the layers
might further facilitate the formation of excitonic insulator,
which would then be reminiscent of the
phenomenon of magnetic catalysis introduced in the abstract field-theoretical setting 
\cite{Miransky}. This scenario was discussed 
\cite{Gorbar,DVK2} in connection with the recent reports of a magnetic field-induced 
insulating behavior in highly oriented pyrolytic graphite (HOPG) \cite{Kopelevich1}.

In the present paper, we numerically solve the gap equation for the excitonic order parameter 
derived in Refs.\cite{DVK1,Gorbar}, thus putting to the test some of the
predictions made in the earlier analytical work on the subject.
 
We start out by reviewing the derivation of the gap equation. 
In the vicinity of the two (in the absence of a lattice strain, 
exactly degenerate and labeled as $i=1,2$) conical $K$-points of the 2D Brillouin zone
of graphite, the low-energy excitations of the valence and the conduction bands 
with a linear dispersion $E_{k}=\pm v_{F}k$ where $v_F\approx 2\times 10^6 m/s$ can
be described as two-component (Weyl) spinors
$\psi_{i\sigma}$, of which there are $N$ different species 
with the spin index $\sigma=1,..., N$ \cite{Semenoff}. 

Provided that the Zeeman coupling
to an external magnetic field is much weaker than the orbital (diamagnetic) term, 
the number of fermion species in graphite is $N=2$, and the dimensionless
Coulomb coupling $g=2\pi e^2/\epsilon v_F$ 
with the dielectric constant $\epsilon\approx 2.8$ appears to be the only relevant parameter.

However, in a strong in-plane field, the Zeeman splitting between the
spin-up and spin-down bands, while having no effect on the 
electron orbital motion, reduces the number of fermions down to $N=1$, which prompts
one to treat $N$ as yet another (to a certain extent) adjustable parameter.

The pair of spinors $\psi_{i\sigma}$
can be further combined into a single  
bi-spinor $\Psi_\sigma=(\psi_{1\sigma}, \psi_{2\sigma})$, thereby allowing one to use 
the Dirac-like representation for a kinetic energy of the two-dimensional electrons   
in a single layer of graphite
\begin{equation}
H_{K}=iv_F\sum_{\sigma =1}^N \int_{\bf r}{\overline \Psi}_\sigma
({\hat \gamma}_1\nabla_x+{\hat \gamma}_2\nabla_y)\Psi_\sigma
\label{kin}
\end{equation}
where ${\overline \Psi}_\sigma=
\Psi^\dagger_\sigma{\hat \gamma}_0$ and the $4\times 4$ (reducible) representation 
of the $\gamma$-matrices ${\hat \gamma}_{0,1,2}=(\tau_{3}, i\tau_{2}, -i\tau_{1})
\otimes\tau_3$ satisfying the anticommutation relations 
$\{{\hat \gamma}_{\mu},{\hat \gamma}_{\nu}\}=2 {\rm diag}(1, -1, -1)$ is 
constructed in terms of the Pauli triplet $\tau_{1,2,3}$.
 
Accordingly, the Coulomb interaction term in the Hamiltonian takes the form 
\begin{equation}
H_C={1\over 4\pi}\sum^{N}_{\sigma,\sigma^\prime=1}\int_{{\bf r},{\bf r}^\prime}
{\overline \Psi}_{\sigma}({\bf r}){\hat \gamma}_0\Psi_{\sigma}
({\bf r}){g\over {|{\bf r}-{\bf r}^\prime|}}
{\overline \Psi}_{\sigma^\prime}({\bf r}^\prime)
{\hat \gamma}_0\Psi_{\sigma^\prime}({\bf r}^\prime)
\label{int}
\end{equation} 
Both Eqs.(1) and (2) remain invariant under arbitrary $U(2N)$ rotations
of the $2N$-component vector $(\Psi_{L\sigma}, \Psi_{R\sigma})$
composed of the chiral $(L,R)$ parts of the Dirac fermion $\Psi_\sigma$ 
defined as $\Psi_{(L,R)\sigma}={1\over 2}({\bf 1}\pm{\hat \gamma}_5)\Psi_\sigma$
where the matrix ${\hat \gamma}_5=
{\bf 1}\otimes{\tau_2}$ anticommutes with any ${\hat \gamma}_\mu$. 

In the relativistic theories of Refs.\cite{Appelquist88,HY}, 
the standard CSB pattern $U(2N)\to U(N)\otimes U(N)$ is signaled
by the development of a singlet order parameter
$\Delta_s({\bf r})=<\sum_{\sigma}^N{\overline \Psi}_\sigma({\bf r})\Psi_\sigma({\bf r})>$,
which is the type of excitonic pairing that we focus upon below.

The standard Dyson-Schwinger equation for the Dirac propagator 
reads (hereafter $p_\mu=(\epsilon, {\bf p})$)
\begin{eqnarray}
\hat{G}^{-1}(\omega, {\bf p})=\hat{G}_{0}^{-1}(\omega, {\bf p}) -   \nonumber\\
T\sum_\Omega\int\frac{d^2{\bf k}}{(2\pi)^{2}}V(\Omega-\omega, {\bf k}-{\bf p})
{\hat \gamma}_{0}\hat{G}(\Omega, {\bf k}){\hat \gamma}_0
\label{eq:schwin}
\end{eqnarray}
where ${\hat G}_0(\omega, {\bf p})={\hat \gamma}_\mu p_\mu/p^2$ is the bare propagator
of the gapless Dirac fermions and the renormalized Coulomb interaction 
\begin{equation}
V(\Omega, {\bf q})=\frac{g}{{\bf q}+Ng\chi(\Omega,{\bf q})}
\label{eq:pot} 
\end{equation}
is modified, as compared to its bare form,
by to the fermion polarization $\chi(\Omega, {\bf q})$.

Being primarily concerned with the possibility of a spontaneous 
gapping of the conical spectrum, we search for a solution to Eq.(\ref{eq:schwin}) 
in the form $\hat{G}(p)=({\hat \gamma}_\mu p_\mu + \Delta(p))^{-1}$,
while neglecting both the vertex and the fermion wave function renormalization.
On the basis of
the experience gained from the relativistic theories \cite{Appelquist88,Aitchison},
one might expect that, albeit being capable of affecting such practically important details
as the value of the critical coupling $g_c(N)$, the latter do not
alter the very existence of a solution (or a lack thereof).

Proceeding along the lines of the previous numerical analyses
of the finite temperature version of $QED_{2+1}$ \cite{Aitchison},
below we focus on the momentum dependence of the static $(\omega=0)$ component
of the order parameter. To this end, we neglect
all but the $\Omega=0$ harmonics of the effective interaction (4) which,
as shown in Refs.\cite{Aitchison}, suffices for determining the critical 
conditions for the emergent CSB order.
In the case of interest, this static approximation appears to be even better justified, 
since the Lorentz invariance is broken already at zero temperature. 

The static component of the finite-temperature fermion polarization 
can, in turn, be approximated by the expression
\begin{equation}
\chi(0,{\bf q})=\frac{1}{8}\:
\left(\left|\mathbf{q}\right|+{T\over C}\exp
\left(-C\frac{\left|\mathbf{q}\right|}{T}\right)\right).
\label{eq:pol}
\end{equation}
which, for $C=\pi/16\ln 2$, provides an up to a few percent 
accurate interpolation between the two opposite limits:
$v_F{\bf q}\gg T$ where Eq.(\ref{eq:pol}) agrees with the zero-temperature 
result, $\chi(0, {\bf q})\sim |{\bf q}|$,
and $v_F{\bf q}\ll T$ where it accounts for thermal 
screening, $\chi(0,{\bf 0})\sim T$  \cite{Aitchison}.

Taking the sum over the Matsubara frequencies in Eq.(3), we then arrive at the gap
equation derived in Refs.\cite{DVK1,Gorbar} (hereafter we use the units $v={\hbar}=1$)
\begin{equation}
\Delta({\bf p}) = \int\frac{d^2{\bf k}}{8\pi^{2}}V(0,{\bf k}-{\bf p})
\Delta({\bf k})
{\tanh{\sqrt{{\bf k}^2+\Delta({\bf k})^2 }\over 2T}\over
{\sqrt{{\bf k}^2+\Delta({\bf k})^{2} }}}
\label{eq:gap}
\end{equation}
This form of the gap equation is more familiar in condensed matter-related applications
where the gap is routinely considered to be a function of momentum but not energy,
consistent with our use of the static approximation.

\begin{figure}[b]
\vspace{0.025in}
\begin{center}
\includegraphics[width=3.0in]{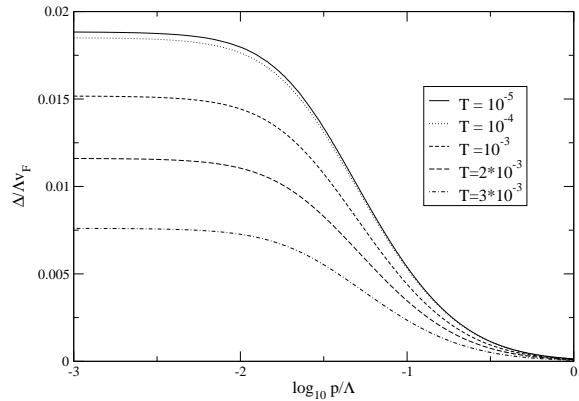}
\end{center}
\caption{Momentum dependence of the solution to the gap equation
 (6) for $N=1$ and different temperatures.}
  \label{temp}
\end{figure}

\begin{figure}[t!]
\vspace{0.05in}
\begin{center}
\includegraphics[width=3.0in]{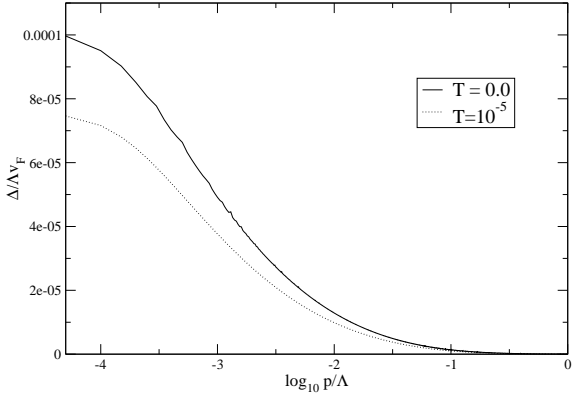}
\end{center}
\caption{Solution to the gap equation for $N=2$.}
  \label{temp2}
\end{figure}

In fact, even a partial account of the momentum and/or energy dependence 
of the gap function goes well beyond the customary BCS-like (constant) solution for the gap.
In the case of short-ranged repulsive interactions, a BCS-like solution of the analog
of Eq.(6) was recently discussed
in conjunction with 
the conjectured excitonic instability in hexaborides \cite{Barzykin}.
By contrast, in a degenerate 
semimetal such as graphite the strong momentum dependence of the unscreened Coulomb interaction 
rules out a constant solution ($\Delta^{BCS}({\bf p})=\Delta$) altogether.

In Fig.1, we present the results of our numerical solution to
Eq.(\ref{eq:gap}) for $N=1$ and several different temperatures.
As a function of momentum, the gap $\Delta({\bf p})$
levels off at ${\bf p}\sim \Delta(0)$, in accord with the approximate analytical solution of 
Refs.\cite{DVK1,Gorbar} where Eq.(\ref{eq:gap}) was substituted with a differential
equation complemented by the boundary conditions $d\Delta({\bf p})/d{\bf p}|_{{\bf p}=0}=0$
and $\Delta({\bf p})+{\bf p}d\Delta({\bf p})/d{\bf p}|_{{\bf p}=\Lambda}=0$,
$\Lambda$ being the upper momentum cutoff given by the maximum span of the Brillouin zone. 

The functional dependence of the solution for $N=2$
is similar to the $N=1$ case, 
apart from the overall suppression by roughly two orders of magnitude
(see Fig.2).

In the strong coupling zero-temperature ($g\to\infty$ and $T\to 0$) limit, the 
$N$-dependence of the zero-momentum gap shown in Fig.3 can be well 
fitted with the formula
\begin{equation}
\Delta(0)= 1.2v_F\Lambda\exp \left( -1.7\pi/{\sqrt {N_{c}^\infty/N-1}}\right)
\end{equation}
which manifests the
Kosterlitz-Thouless nature of the excitonic transition 
that occurs at $T=0$ and 
$N_{c}^\infty\equiv N_c(g=\infty)\approx 2.6$, in agreement with the predictions
made in Refs.\cite{DVK1,Gorbar}. 
 
\begin{figure}[b!]
\vspace{0.05in}
\begin{center}
\includegraphics[width=3.0in]{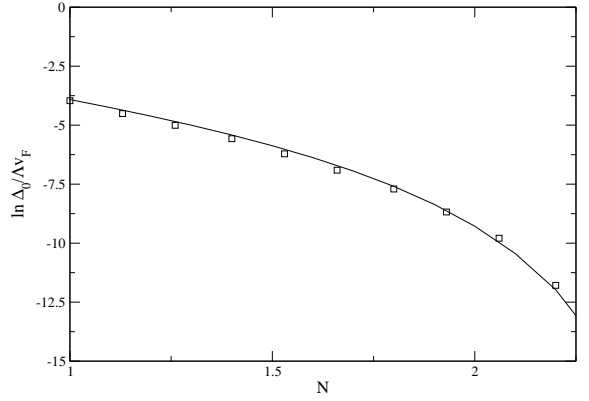}
\end{center}
\vspace{0.05in}
\caption{Zero-momentum gap for different values of $N$ at 
 \(T=10^{-5}\) and \(g\rightarrow \infty\). 
The solid line is the fit given by Eq.(7).}
  \label{kos}
\end{figure}

In Fig.4, we present the $N$-dependence of the critical coupling $g_{c}(N)$.
For $N=2$, our numerical result $g_c(N=2)\approx 7$ differs by a factor
of two from that obtained analytically in Ref.\cite{Gorbar}, although  
the agreement improves for smaller $N$.

Notably, the relevant values of
$g_c$ are rather large, which indicates that any weak-coupling approach
would be utterly inadequate for the problem in question.

Even with the possible caveat that the gap equation 
tends to systematically underestimate the critical strength of the repulsive interactions
\cite{Appelquist99,Hands}, the above values compare 
favorably with the estimates $g\sim 5-10$ obtained for the 
HOPG samples of Ref.\cite{Kopelevich1}.

\begin{figure}[t!]
\vspace{0.05in}
\begin{center}
\includegraphics[width=3.0in]{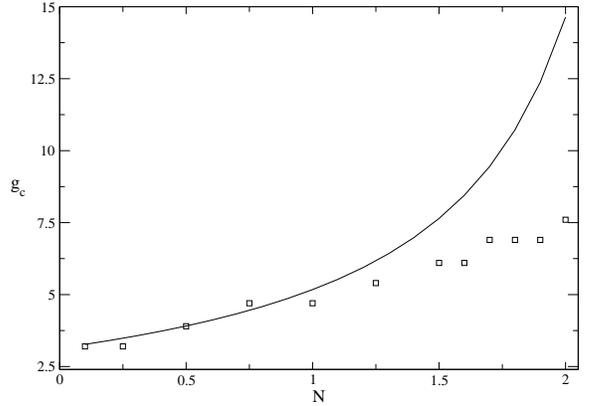}
\end{center}
\vspace{0.05in}
\caption{The critical coupling $g_{c}$ for different values of $N$ and $T=0$.
The solid line shows the analytical result of Ref.10.}
  \label{g}
\end{figure}

For any $N$, the zero-momentum gap attains its maximum value $\Delta(0)\approx
10^{-4}v_F\Lambda$
at $g\to\infty$.
Using the parameters of the HOPG band structure, 
we estimate the maximal possible gap at $N=2$
as $\Delta(0)\sim 30K$, which turns out to be small compared to the 
typical Fermi energy $E_F\sim 250K$ in HOPG \cite{Kopelevich1}. The latter 
characterizes the actual (as opposed to the idealized point-like)
Fermi surface of graphite which represents
a combined effect of inter-layer hopping, finite doping, and/or disorder.

At this point, it remains unknown to what extent the above factors
can modify Eq.(6) derived for a clean two-dimensional sheet of undoped graphite.
On the other hand, in a layered system the propensity 
towards excitonic pairing is further strengthened by 
the inter-layer Coulomb repulsion, which, 
considering the fact that $g>1$, might be even more important than the finite $E_F$ \cite{DVK1}.

To this end, it is worth mentioning that the singlet excitonic
order parameter is directly related to the electron density 
imbalance between the $A$ and $B$ sublattices 
of the bi-partite hexagonal lattice: $\Delta_s({\bf p}={\bf 0}, {\bf r})=
\sum_{\sigma=1,2}<{\overline \Psi}_\sigma({\bf r}) \Psi_\sigma({\bf r})>
=\sum_{i\sigma=1,2}(\delta_{{\bf r}, A}<\psi^\dagger_{i\sigma}(A)\psi_{i\sigma}(A)>-
\delta_{{\bf r}, B}<\psi^\dagger_{i\sigma}(B)\psi_{i\sigma}(B)>)$.
 
In a multi-layer system stacked in the staggered ($A-B$) configuration,
the inter-layer Coulomb repulsion favors spontaneous depletion (respectively, pile-up)
of the electron density on a sublattice formed by the carbon 
atoms above and below
the centers (respectively, corners) of the hexagons in adjacent layers. 
The resulting commensurate charge density wave
alternates between the layers, thereby keeping the electrons in the adjacent layers 
as far apart as possible and nudging the system closer to the excitonic instability. 

In order to decide on
the ultimate outcome of the competition between the frustrating effect of the extended Fermi surface
and the strong (yet, non-singular, unlike Eq.(2)) 
inter-layer Coulomb repulsion, the present analysis has to be further refined
by incorporating the above factors through, e.g., a non-linear
fermion dispersion and effective four-fermion terms \cite{Helio2}. 

Although, thus far, we were only interested in the possibility of singlet pairing,
for $N=2$ and in the leading Coulomb approximation (small transferred momenta)
the formation of a triplet order parameter appears to be just as likely.
In fact, similar to the case of the short-ranged repulsion
discussed in Refs.\cite{Balentz}, the 
excitonic ground state possesses a degeneracy
with respect to arbitrary $SU(4)$ rotations of the 
four-dimensional complex vector composed of 
the singlet $\Delta_s$ and the triplet ${\vec \Delta}_t=\sum_{\sigma,\sigma^\prime=1,2}
<{\overline \Psi}_\sigma{\vec \sigma}_{\sigma\sigma^\prime}\Psi_{\sigma^\prime}>$ 
order parameters. 

This approximate degeneracy gets lifted upon including the (so far, neglected) 
short-ranged Coulomb exchange interactions 
which cause transitions between the conduction and valence bands \cite{Balentz}. 
Alongside the Zeeman coupling, the latter favor the triplet order parameter, 
thus enforcing the Hund's rule.

It is also conceivable that, with increasing doping,
the triplet excitonic insulator 
can give way to an itinerant ferromagnetic
metal. The interest in this scenario which had been previously
discussed only in the case of the three-dimensional non-degenerate semimetals
was bolstered by the discovery of a weak ($\sim 0.07\mu_B$ per
carrier), yet robust ($T_c\sim 600-1000 K$), ferromagnetism in hexaborides \cite{Barzykin}. 

However, more recent experimental studies of hexaborides \cite{exp-hexaborides}
have indicated
the presence of a large spectral gap which seems to rule out the excitonic mechanism, 
thus putting graphite in a rather unique position of the best
currently known candidate for excitonic ferromagnet.
This possibility appears to be particularly intriguing in the light of the reports
of a comparably weak ($\sim 0.1\mu_B$ per carrier) magnetization observed in
HOPG samples at room temperatures \cite{Kopelevich2}.

The conditions for the emergence of excitonic 
ferromagnetism and the putative global phase diagram of 
graphite will be discussed at a greater length in our future work \cite{Helio2}.

To summarize, in the present paper we obtained a numerical solution to the gap equation
describing the conjectured excitonic transition in two-dimensional
semimetals such as graphite. From our solution, we inferred
the minimal necessary strength of the 
Coulomb coupling $g_{c}$ and the largest possible number of fermion species
$N_{c}^\infty$
for which this Kosterlitz-Thouless-type transition may occur at zero temperature.
Although this analysis was carried out for a single undoped layer, 
we believe that our predictions for the existence of a critical Coulomb coupling, 
momentum dependence of the gap function, and the nature of the excitonic transition
should remain robust upon including further complicating
factors such as inter-layer coupling, doping, and disorder.

The authors gratefully acknowledge the North Carolina Supercomputing Center 
for providing access to their computing facilities.
This research was supported by the NSF under Grant DMR-0071362.

\wt

\begin{references}

\bibitem{DVK1}
D.V.Khveshchenko, Phys. Rev. Lett. {\bf 87}, 246802 (2001).

\bibitem{Appelquist88}
T.Appelquist, D.Nash, and L.C.R. Wijewardhana, 
Phys.Rev.Lett.{\bf 60}, 2575 (1988);
T.Appelquist, J.Terning, and L.C.R. Wijewardhana, 
ibid {\bf 75}, 2081 (1995).

\bibitem{HY} B. Rosenstein, B.J. Warr, and S.H. Park, Phys. Rep. {\bf 205}, 59 (1991);
S. Hands, A. Koci\'{c}, and J.B. Kogut, Annals of Phys.{\bf 224}, 29 (1993).

\bibitem{RW} D.H. Kim and P.A. Lee, Annals of Phys. {\bf 272}, 133 (1999);
W. Rantner and X.-G. Wen, Phys. Rev. Lett. {\bf 86} (2001) 3871; cond-mat/0105540,0201521. 

\bibitem{FT} M. Franz, Z. Tesanovic, and O. Vafek, Phys. Rev. {\bf B66}, 054535 (2002); 
J. Ye, Phys. Rev.{\bf B65}, 214505 (2002). 
\bibitem{Herbut} I. F. Herbut, Phys. Rev. Lett. {\bf 88} (2002) 047006;
Phys. Rev.{\bf B66}, 094504 (2002); B. H. Seradjeh and I. F. Herbut, ibid {\bf B 66}, 184507 (2002).

\bibitem{Appelquist99} 
T. Appelquist, A. G. Cohen, and M. Schmaltz, 
Phys. Rev. {\bf D60}, 045003 (1999).

\bibitem{Hands} S. J. Hands, J. B. Kogut, C. G. Strouthos, 
Nucl. Phys. {\bf B645}, 321 (2002); hep-lat/0209133.

\bibitem{Lee} D. Lee, cond-mat/0212204.

\bibitem{Gorbar}
E. V. Gorbar, V. P.Gusynin, V. A. Miransky, and I. A. Shovkovy, Phys. Rev. {\bf B66}, 045108 (2002).

\bibitem{DVK2}
D. V. Khveshchenko, Phys. Rev. Lett. {\bf 87}, 206401 (2001).

\bibitem{Miransky}
V. P. Gusynin, V. A. Miransky, and I. A. Shovkovy, 
Phys. Rev. Lett. {\bf 73}, 3499 (1994);
Phys. Rev. {\bf D52}, 4718 (1995);
Phys. Rev. {\bf D61}, 045005 (2000);
J. Alexandre, K. Farakos, and G. Koutsoumbas, 
ibid {\bf D62}, 105017 (2000); 
ibid {\bf D63}, 065015 (2001).

\bibitem{Kopelevich1}
H. Kempa, P. Esquinazi, and Y. Kopelevich, Phys. Rev. {\bf B 65}, 241101(R) (2002);
M. S. Sercheli et al, Solid State Comm. {\bf 121}, 579 (2002);
H. Kempa et al, ibid {\bf 125} 1 (2003).

\bibitem{Semenoff} 
G. Semenoff, Phys.Rev.Lett.{\bf 53}, 2449 (1984);
F. D. M. Haldane, ibid {\bf 61}, 2015 (1988); 
J. Gonzalez, F. Guinea, and M. A. H. Vozmediano, ibid {\bf 69}, 172 (1992);
Nucl.Phys.{\bf 406}, 771 (1993).

\bibitem{Aitchison} 
I. J. R. Aitchison et al, Phys.Lett.{\bf B294}, 91 (1992); 
N. Dorey and N. Mavromatos, Nucl. Phys.{\bf B386}, 614 (1992);
I. J. R. Aitchison and M. Klein-Kreisler, Phys.Rev.{\bf D50}, 1068 (1994);
I. J. R. Aitchison, Z. Phys.{\bf C67}, 303 (1995);
I. J. R. Aitchison and N. Mavromatos, Nucl.Phys.{\bf B}, 614 (1996);
G. Triantaphyllou, Phys. Rev.{\bf D58}, 065006 (1998).

\bibitem{Barzykin}
D. P. Young et al, Nature {\bf 397}, 412 (1999);
M. E. Zhitomirsky, T. M. Rice, and V. I. Anisimov, Nature (1999);
L. Balentz and C. M. Varma, Phys. Rev. Lett.{\bf 84}, 1246 (2000);
V. Barzykin and L. P. Gorkov, ibid {\bf 84}, 2207 (2000).

\bibitem{Helio2} H. Leal and D. V. Khveshchenko, unpublished.

\bibitem{Balentz}
L. Balents, Phys. Rev.{\bf B62}, 2346 (2000); 
M. Y. Veillette and L. Balents, ibid {\bf B65}, 014428 (2002); 
E. Bascones, A. A. Burkov, and A. H. MacDonald, 
Phys. Rev. Lett. {\bf 89}, 086401 (2002). 

\bibitem{exp-hexaborides}
J. D. Denlinger et al,
Surf. Rev. Lett. {\bf 9}, 1309 (2002); 
Phys. Rev. Lett. {\bf 89}, 157601 (2002).

\bibitem{Kopelevich2}
Y. Kopelevich et al, J. Low Temp. Phys.{\bf 119}, 691 (2000);
P. Esquinazi et al,
Phys. Rev. {\bf B66}, 024429 (2002);
Y. Kopelevich et al, cond-mat/0209442.



\end{references}
\end{document}